\documentclass[12pt,preprint2]{aastex6}

\usepackage{graphicx}				
\usepackage{hyperref}
\usepackage{natbib}

\newcommand\msun{M$_{\odot}$}
\newcommand\ocen{$\omega$ Cen}
\newcommand\ocent{$\omega$ Centauri}

\newcommand{\teff}{$T_{\rm eff}$} 
\newcommand{\logg}{$\log g$} 
\newcommand{\kms}{km s$^{-1}$}
\newcommand{\vt}{$\xi_t$} 
\newcommand{\fei}{Fe\,{\sc i}}

\newcommand{\pbi}{Pb\,{\sc i}}
\newcommand{\cui}{Cu\,{\sc i}}
\newcommand{\yii}{Y\,{\sc ii}}

\newcommand{\baii}{Ba\,{\sc ii}}

\begin{document}
\title{A chemical signature from fast-rotating low-metallicity massive stars:
ROA 276 in $\omega$ Centauri\footnote{This paper includes data gathered with the 6.5
meter Magellan Telescopes located at Las Campanas Observatory, Chile.}}

\author{David Yong\altaffilmark{1}, 
John E.\ Norris\altaffilmark{1}, 
Gary S.\ Da Costa\altaffilmark{1}, 
Laura M.\ Stanford\altaffilmark{1}, 
Amanda I.\ Karakas\altaffilmark{2,1}, 
Luke J.\ Shingles\altaffilmark{3}, 
Raphael Hirschi\altaffilmark{4,5}, 
Marco Pignatari\altaffilmark{6,7,8} }
\altaffiltext{1}{Research School of Astronomy and Astrophysics, Australian National
University, Canberra, ACT 2611, Australia} 
\altaffiltext{2}{Monash Centre for Astrophysics, School of Physics \& Astronomy, Monash
University, Victoria 3800, Australia} 
\altaffiltext{3}{Astrophysics Research Centre, School of Mathematics and Physics, Queen's
University Belfast, Belfast BT7 1NN, UK}
\altaffiltext{4}{Astrophysics Group, Keele University, Staffordshire ST5 5BG,
UK} 
\altaffiltext{5}{Kavli Institute for the Physics and Mathematics of the Universe (WPI),
The University of Tokyo, Kashiwa, Chiba 277-8583, Japan} 
\altaffiltext{6}{E.A. Milne Centre for Astrophysics, Department of Physics \& Mathematics,
University of Hull, HU6 7RX, UK}
\altaffiltext{7}{Konkoly Observatory, Hungarian Academy of Sciences, Budapest,
Hungary} 
\altaffiltext{8}{NuGrid collaboration http://www.nugridstars.org}

\begin{abstract}

We present a chemical abundance analysis of a metal-poor star, ROA 276, in the
stellar system \ocent. We confirm that this star has an unusually high [Sr/Ba]
abundance ratio. Additionally, ROA 276 exhibits remarkably high abundance
ratios, [X/Fe], for all elements from Cu to Mo along with normal abundance
ratios for the elements from Ba to Pb. The chemical abundance pattern of ROA
276, relative to a primordial \ocen\ star ROA 46, is best fit by a
fast-rotating low-metallicity massive stellar model of 20 \msun, [Fe/H] =
$-$1.8, and an initial rotation 0.4 times the critical value; no other
nucleosynthetic source can match the neutron-capture element distribution. ROA
276 arguably offers the most definitive proof to date that fast-rotating
massive stars contributed to the production of heavy elements in the early
Universe. 

\end{abstract}


\section{Introduction} \label{sec:intro}

Numerical simulations predict that low-metallicity stars that formed in the
early Universe were massive, compact, and rotated near their critical
velocities where gravity is balanced by centrifugal forces
\citep{Bromm:2004aa,Stacy:2011aa}. Nucleosynthesis in these fast-rotating
low-metallicity massive stars (hereafter spinstars) differs considerably from
their non-rapidly-rotating counterparts
\citep{Meynet:2006aa,Hirschi:2007aa,Pignatari:2008aa,Frischknecht:2012aa,Frischknecht:2016aa,Maeder:2012aa}. 
Since these massive stars have long since died, confirmation of their existence
can be obtained by identifying their unique chemical signature in the abundance
patterns of subsequent generations of Milky Way stars
\citep{Frebel:2015aa,Maeder:2015aa}. 

One chemical signature of spinstars comes from nitrogen abundances in
metal-poor halo stars which require primary production \citep{Spite:2005aa}. 
While spinstars can naturally achieve such nucleosynthesis, hydrogen ingestion
in massive stars \citep{Pignatari:2015aa} and intermediate-mass and super
asymptotic giant branch (AGB) stars \citep{Karakas:2010aa,Doherty:2014aa} may
also be responsible for nitrogen production in the early Universe. 

Another possible observational signature of spinstars comes from
neutron-capture elements. The scatter in Sr and Ba abundances in
low-metallicity halo stars can be explained by spinstars
\citep{Cescutti:2013aa}, but measurements of other neutron-capture elements
(e.g., Y, Zr, La), when available, are also compatible with massive AGB stars
\citep{Fishlock:2014aa}. \citet{Chiappini:2011aa} reported unusually high
abundances for the elements Sr, Y, Ba, and La in the bulge globular cluster NGC
6522, consistent with yields from spinstars. Those measurements, however, have
since been revised downwards and could also be explained by AGB stars
\citep{Barbuy:2014aa,Ness:2014aa}. The unmistakable signature among the 
neutron-capture elements from spinstars has yet to be seen within an individual
star. 

\section{Target Selection and Observations} \label{sec:obs}

\ocent\ is the most massive star cluster in our Galaxy. In contrast to the
majority of Milky Way globular clusters, \ocen\ exhibits a number of peculiar
features including a broad range in abundances for iron and slow
neutron-capture process, or $s$-process, elements \citep{Norris:1995aa}. The
distribution and evolution of the $s$-process element abundances in \ocen\ are
consistent with a dominant contribution from 1.5 - 3 \msun\ AGB stars
\citep{Smith:2000aa}. 

There are two stars in \ocen, however, that exhibit peculiar abundance ratios
of Sr and Ba \citep{Stanford:2006ab,Stanford:2010aa}; the red giant ROA 276
with $V$ = 12.37 and the main sequence star 2015448 with $V$ = 18.22. Both
objects have high Sr and low Ba abundances, consistent with predictions of
neutron-capture nucleosynthesis in spinstars
\citep{Frischknecht:2012aa,Frischknecht:2016aa}. 

To further examine these unusual abundance patterns, we obtained a
high-resolution optical spectrum for the red giant ROA 276 and a comparison
star ROA 46 ($V$ = 11.54) using the Magellan Inamori Kyocera Echelle
spectrograph \citep{Bernstein:2003aa} at the 6.5m Magellan Clay Telescope on
2007 June 22-23. Both stars have proper motions and radial velocities
consistent with cluster membership \citep{Bellini:2009ab}. The total exposure
time was 10 min per target. We used the 0.5\arcsec\ slit to achieve a spectral
resolution of $R$ = 56,000 and $R$ = 44,000 in the blue and red arms,
respectively. One dimensional, wavelength calibrated, continuum normalized
spectra were produced from the raw spectra using {\sc iraf}\footnote{{\sc IRAF}
is distributed by the National Optical Astronomy Observatory, which is operated
by the Association of Universities for Research in Astronomy (AURA) under a
cooperative agreement with the National Science Foundation.} and the {\sc
mtools}\footnote{\url{www.lco.cl/telescopes-information/magellan/instruments/mike/iraf-tools/iraf-mtools-package}}
package. The signal-to-noise ratio (S/N) for both stars was roughly 80 per
pixel near 6000\,\AA\ and 40 per pixel near 4500\,\AA. The spectra have
approximately 3.5 pixels per resolution element. 

\section{Stellar Parameters and Chemical Abundances} \label{sec:params}

The stellar parameters were determined from a traditional spectroscopic
approach following the procedure outlined in \citet{Yong:2014aa}. Equivalent
widths (EWs) were measured using routines in {\sc iraf} and {\sc daospec}
\citep{Stetson:2008aa}, and there was good agreement between the two
approaches. Weak (EW $<$ 10\,m\AA) and strong (EW $>$ 150\,m\AA) lines were
removed from the analysis. Abundances were derived using the EW,
one-dimensional local thermodynamic equilibrium (LTE) model atmospheres with
[$\alpha$/Fe] = +0.4 \citep{Castelli:2003aa}, and the LTE stellar line analysis
program {\sc moog} \citep{Sneden:1973aa}. The version of {\sc moog} that
we used includes a proper treatment of Rayleigh scattering
\citep{Sobeck:2011aa}. The effective temperature (\teff), surface gravity
(\logg), and microturbulent velocity (\vt), were obtained by enforcing
excitation and ionization balance for Fe lines (see Table \ref{tab:param}). The
uncertainties in \teff, \logg, and \vt\ are 50\,K, 0.2\,dex, and 0.2\,\kms,
respectively. The standard deviation for \fei\ lines was 0.19 dex (ROA 276) and
0.16 dex (ROA 46), and we adopted an uncertainty in the model atmosphere of
[m/H] = 0.2 dex. 

\begin{deluxetable}{l|cccc|cc}[h!]
\tablecaption{Stellar parameters for the program stars. \label{tab:param}} 
\tablecolumns{7}
\tablewidth{0pt}
\tablehead{
\colhead{Star} &
\colhead{\teff} &
\colhead{\logg} &
\colhead{\vt} &
\colhead{[Fe/H]} &
\colhead{\teff} &
\colhead{\logg} \\ 
\colhead{} &
\colhead{(K)} &
\colhead{(cgs)} &
\colhead{(\kms)} &
\colhead{(dex)} &
\colhead{(K)} &
\colhead{(cgs)} \\ 
\colhead{} &
\multicolumn{4}{c}{Spectroscopic} & 
\multicolumn{2}{c}{Photometric} 
}
\startdata
ROA 276	& 4125 & 0.70 & 1.75 & $-$1.30 & 4130 & 0.79 \\
ROA 46  & 4075 & 0.20 & 2.40 & $-$1.72 & 4024 & 0.37 \\ 
\enddata
\end{deluxetable}

Stellar parameters can also be derived from a photometric approach. \teff\ can
be estimated from color-temperature relations based upon the infrared flux
method \citep{Blackwell:1977aa,Ramirez:2005ab}. We used $BVRIJHK$ photometry
\citep{Bellini:2009ab,Skrutskie:2006aa} and adopted a reddening of $E(B-V)$ =
0.12 \citep[][2010 edition]{Harris:1996aa}. The surface gravity can be
determined assuming the photometric \teff, a distance modulus $(m-M)_V$ = 13.94
\citep[][2010 edition]{Harris:1996aa}, bolometric corrections from
\citet{Alonso:1999aa}, and a mass of 0.8 \msun. \teff\ and \logg\ obtained from
the spectroscopic and photometric approaches are in good agreement when
considering the estimated uncertainties (see Table \ref{tab:param}). 

\begin{figure*}
\centering
\includegraphics[width=4.5in]{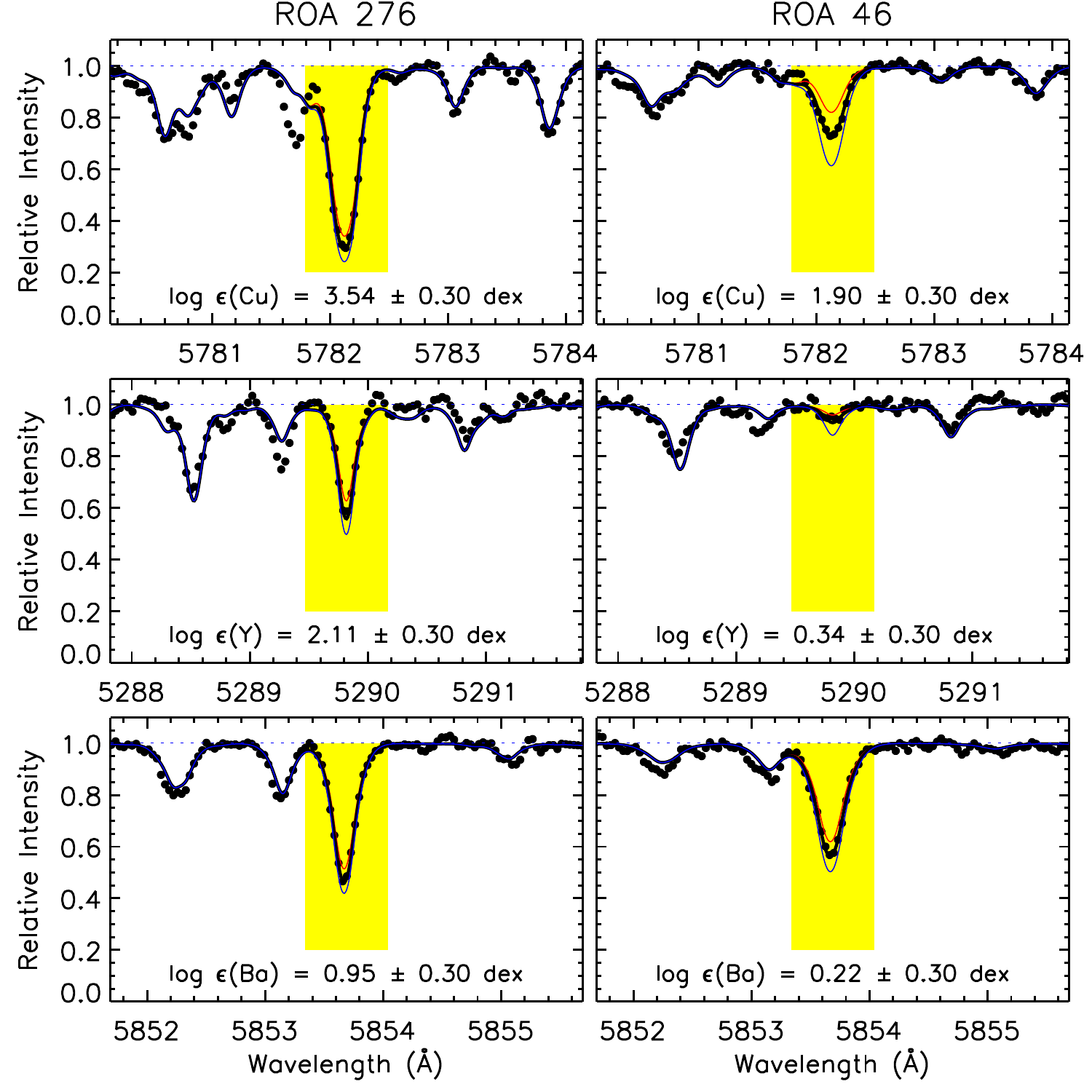}
\caption{Observed and synthetic spectra for ROA 276 (left panels) and ROA 46
(right panels) for some representative elements. From top to bottom, the
spectral lines are \cui\ 5782.14\,\AA, \yii\ 5289.82\,\AA, 
and \baii\ 5853.69\,\AA. Filled circles represent the observed spectra. The
thick black line is the best fitting synthetic spectra and unsatisfactory fits
($\pm$ 0.3 dex) are included as thin red and blue lines. In each panel, we
write the final abundance and the region within which the $\chi^2$ minimization
was computed is indicated in yellow.} 
\label{fig:spec}
\end{figure*}

Elemental abundances were derived using {\sc moog} for individual lines based
on the EW or from spectrum synthesis following \citet{Yong:2014aa}. Examples of
synthetic spectra fits for representative lines of selected elements are given
in Figure \ref{fig:spec}. Aside from the 4057.81\,\AA\ \pbi\ line, given the
S/N of the blue spectra we analyzed lines redward of 4317.31\,\AA. We present
our line list, EWs, and abundance measurements in Table \ref{tab:line}. Solar
abundances were taken from \citet{Asplund:2009aa} and the sources of the $gf$
values can be found in Table \ref{tab:line}. 

\begin{deluxetable*}{lccrccccc}
\tablecaption{Line list, equivalent widths, and abundances. \label{tab:line}} 
\tablecolumns{9}
\tablewidth{0pt}
\tablehead{
\colhead{Species} &
\colhead{Wavelength} &
\colhead{LEP} &
\colhead{$\log gf$} &
\colhead{EW (ROA 276)} &
\colhead{EW (ROA 46)} &
\colhead{$\log\epsilon$(X) (ROA 276)} & 
\colhead{$\log\epsilon$(X) (ROA 46)} & 
\colhead{Source} \\ 
\colhead{} &
\colhead{(\AA)} &
\colhead{(eV)} &
\colhead{} &
\colhead{(m\AA)} &
\colhead{(m\AA)} &
\colhead{(dex)} & 
\colhead{(dex)} & 
\colhead{} 
}
\startdata
CH    & 4270 - 4330 &   &          &      syn &      syn &     6.68 &     6.08 &    1 \\ 
O  I  & 6300.31 &  0.00 &  $-$9.75 &     69.6 &     69.7 &     8.00 &     7.60 &    2 \\	
O  I  & 6363.78 &  0.02 & $-$10.25 &     36.9 &   \ldots &     8.09 &   \ldots &    3 \\ 
Na I  & 5682.65 &  2.10 &  $-$0.67 &     64.8 &   \ldots &     4.82 &   \ldots &    2 \\
Na I  & 5688.22 &  2.10 &  $-$0.37 &     87.4 &     68.7 &     4.86 &     4.55 &    2 \\
\enddata
\tablerefs{
(1) = \citet{Masseron:2014aa};
(2) = \citet{Gratton:2003aa} and references therein; 
(3) = values as used in \citet{Yong:2005aa} where the references include \citet{Kurucz:1995aa}, \citet{Prochaska:2000ab}, \citet{Den-Hartog:2003aa}, \citet{Ivans:2001aa}, and \citet{Ramirez:2002aa}; 
(4) = Oxford group including \citet{Blackwell:1979aa}, \citet{Blackwell:1979ab}, \citet{Blackwell:1980aa}, \citet{Blackwell:1986aa}, \citet{Blackwell:1995aa}; 
(5) = \citet{Kock:1968aa};
(6) = \citet{Hannaford:1983aa}; 
(7) = \citet{Roederer:2012ac}; 
(8) = mean of lifetimes from \citet{Simsarian:1998aa} and \citet{Volz:1996aa} weighted according to uncertainties, via \citet{Morton:2000aa}; 
(9) = \citet{Fuhr:2009aa}; 
(10) = \citet{Biemont:2011aa}; 
(11) = \citet{Biemont:1981aa}; 
(12) = \citet{Ljung:2006aa}; 
(13) = \citet{Whaling:1988aa}; 
(14) = \citet{Davidson:1992aa} using hfs/IS from \citet{McWilliam:1998aa}; 
(15) = \citet{Lawler:2001aa}, using hfs from \citet{Ivans:2006aa}; 
(16) = \citet{Lawler:2009aa}; 
(17) = \citet{Li:2007aa}; 
(18) = \citet{Den-Hartog:2003aa}, using hfs/IS from \citet{Roederer:2008aa} when available;
(19) = \citet{Lawler:2006aa}, using hfs/IS from \citet{Roederer:2008aa} when available;
(20) = \citet{Lawler:2001ab}, using hfs/IS from \citet{Ivans:2006aa};
(21) = \citet{Biemont:2000aa}, using hfs/IS from \citet{Roederer:2012ab}. 
}
\tablecomments{Table \ref{tab:line} is published in its entirety in the
machine-readable format. A portion is shown here for guidance regarding its
form and content.}
\end{deluxetable*}

Uncertainties in chemical abundances were obtained by repeating the analysis
and varying the stellar parameters, one at a time, by their uncertainties.
These four error terms were added, in quadrature, to obtain the systematic
uncertainty. We replaced the random error (s.e.$_{\log \epsilon}$)
by max(s.e.$_{\log \epsilon}$, 0.20/$\sqrt{N_{\rm lines}}$) where the second
term is what would be expected for a set of $N_{\rm lines}$ with a dispersion
of 0.20 dex. The total error was obtained by adding the random and systematic
errors in quadrature. Chemical abundances and their errors are presented in
Table \ref{tab:abun}. 

\begin{table*}[h!]
\centering
\caption{Chemical abundances of ROA 276 and the comparison star ROA 46.}\label{tab:abun}
\begin{tabular}{c|DDDD|DDDD|D}
\tablewidth{0pt}
\decimals
\hline
\hline
Species & 
\multicolumn2c{$\log\epsilon$} & 
\multicolumn2c{$\sigma_{\log\epsilon}$} & 
\multicolumn2c{[Fe/H]} & 
\multicolumn2c{$\sigma_{\rm [Fe/H]}$} & 
\multicolumn2c{$\log\epsilon$} & 
\multicolumn2c{$\sigma_{\log\epsilon}$} & 
\multicolumn2c{[Fe/H]} & 
\multicolumn2c{$\sigma_{\rm [Fe/H]}$} & 
\multicolumn2c{$\Delta$[Fe/H]} \\
&
\multicolumn{8}{c}{ROA 276} & 
\multicolumn{8}{c}{ROA 46} & 
\multicolumn{2}{c}{ROA 276 $-$} \\ 
&
\multicolumn{8}{c}{} & 
\multicolumn{8}{c}{} & 
\multicolumn{2}{c}{ROA 46} \\ 
\hline
Fe I   &    6.20 &    0.01 & $-$1.30 &    0.08 &    5.78 &    0.02 & $-$1.72 &    0.08 &    0.42 \\
Fe II  &    6.21 &    0.04 & $-$1.29 &    0.15 &    5.80 &    0.03 & $-$1.70 &    0.13 &    0.41 \\
\hline
& 
\multicolumn2c{$\log\epsilon$} & 
\multicolumn2c{$\sigma_{\log\epsilon}$} & 
\multicolumn2c{[X/Fe]} & 
\multicolumn2c{$\sigma_{\rm [X/Fe]}$} & 
\multicolumn2c{$\log\epsilon$} & 
\multicolumn2c{$\sigma_{\log\epsilon}$} & 
\multicolumn2c{[X/Fe]} & 
\multicolumn2c{$\sigma_{\rm [X/Fe]}$} & 
\multicolumn2c{$\Delta$[X/Fe]} \\
&
\multicolumn{8}{c}{ROA 276} & 
\multicolumn{8}{c}{ROA 46} & 
\multicolumn{2}{c}{ROA 276 $-$} \\ 
&
\multicolumn{8}{c}{} & 
\multicolumn{8}{c}{} & 
\multicolumn{2}{c}{ROA 46} \\ 
\hline
C (CH) &    6.68 &    0.20 & $-$0.45 &    0.24 &    6.08 &    0.20 & $-$0.63 &    0.24 &    0.18 \\ 
O I    &    8.04 &    0.05 &    0.65 &    0.18 &    7.60 &    0.20 &    0.63 &    0.22 &    0.02 \\
Na I   &    4.76 &    0.08 & $-$0.19 &    0.13 &    4.55 &    0.20 &    0.03 &    0.21 & $-$0.22 \\
Mg I   &    6.85 &    0.02 &    0.54 &    0.13 &    6.34 &    0.20 &    0.46 &    0.22 &    0.08 \\
Ca I   &    5.54 &    0.03 &    0.50 &    0.14 &    5.00 &    0.04 &    0.38 &    0.13 &    0.12 \\
Sc II  &    1.48 &    0.06 & $-$0.37 &    0.14 &    1.49 &    0.06 &    0.06 &    0.13 & $-$0.43 \\
Ti I   &    4.12 &    0.02 &    0.47 &    0.13 &    3.70 &    0.03 &    0.47 &    0.13 &    0.00 \\
Ti II  &    4.21 &    0.05 &    0.56 &    0.14 &    3.69 &    0.03 &    0.46 &    0.12 &    0.10 \\
Cr I   &    4.45 &    0.06 &    0.11 &    0.11 &    4.03 &    0.12 &    0.10 &    0.16 &    0.01 \\
Cr II  &    4.35 &    0.12 &    0.01 &    0.18 &    4.12 &    0.07 &    0.20 &    0.17 & $-$0.19 \\
Mn I   &    3.71 &    0.03 & $-$0.42 &    0.10 &    3.36 &    0.06 & $-$0.35 &    0.10 & $-$0.07 \\
Co I   &    3.91 &    0.04 &    0.22 &    0.17 &    3.25 &    0.20 & $-$0.02 &    0.21 &    0.24 \\
Ni I   &    5.27 &    0.03 &    0.35 &    0.08 &    4.56 &    0.03 &    0.06 &    0.09 &    0.29 \\
Cu I   &    3.56 &    0.02 &    0.67 &    0.17 &    1.96 &    0.06 & $-$0.51 &    0.16 &    1.18 \\
Zn I   &    4.73 &    0.18 &    1.47 &    0.20 &    2.98 &    0.09 &    0.14 &    0.16 &    1.33 \\
Rb I   &    3.15 &    0.03 &    1.93 &    0.17 &    1.22 &    0.20 &    0.42 &    0.23 &    1.51 \\
Sr I   &    2.90 &    0.03 &    1.32 &    0.17 &    0.78 &    0.20 & $-$0.37 &    0.21 &    1.69 \\
Y II   &    2.21 &    0.08 &    1.30 &    0.14 &    0.33 &    0.08 & $-$0.17 &    0.15 &    1.47 \\
Zr I   &    3.12 &    0.13 &    1.84 &    0.16 &    1.16 &    0.08 &    0.30 &    0.10 &    1.54 \\
Zr II  &    3.02 &    0.41 &    1.74 &    0.43 &    1.29 &    0.06 &    0.43 &    0.17 &    1.31 \\
Mo I   &    1.92 &    0.20 &    1.34 &    0.22 &    0.11 &    0.20 & $-$0.05 &    0.22 &    1.39 \\
Ba II  &    1.21 &    0.09 &    0.33 &    0.14 &    0.56 &    0.12 &    0.09 &    0.14 &    0.24 \\
La II  & $-$0.12 &    0.09 &    0.08 &    0.13 & $-$0.66 &    0.07 & $-$0.04 &    0.10 &    0.12 \\
Ce II  &    0.07 &    0.09 & $-$0.21 &    0.13 & $-$0.32 &    0.07 & $-$0.18 &    0.13 & $-$0.03 \\
Pr II  & $-$0.62 &    0.16 & $-$0.04 &    0.19 & $-$1.25 &    0.00 & $-$0.25 &    0.15 &    0.21 \\
Nd II  &    0.22 &    0.05 &    0.10 &    0.10 & $-$0.26 &    0.04 &    0.04 &    0.07 &    0.06 \\
Sm II  & $-$0.64 &    0.06 & $-$0.30 &    0.13 & $-$0.78 &    0.05 & $-$0.02 &    0.11 & $-$0.28 \\
Eu II  & $-$0.76 &    0.20 &    0.02 &    0.22 & $-$1.17 &    0.20 &    0.03 &    0.20 & $-$0.01 \\
Pb I   &    0.70 &    0.20 &    0.25 &    0.23 &    0.45 &    0.20 &    0.42 &    0.24 & $-$0.17 \\
\hline
\end{tabular}
\end{table*}

\section{Results} \label{sec:results}

\begin{figure}
\centering
\includegraphics[width=\columnwidth]{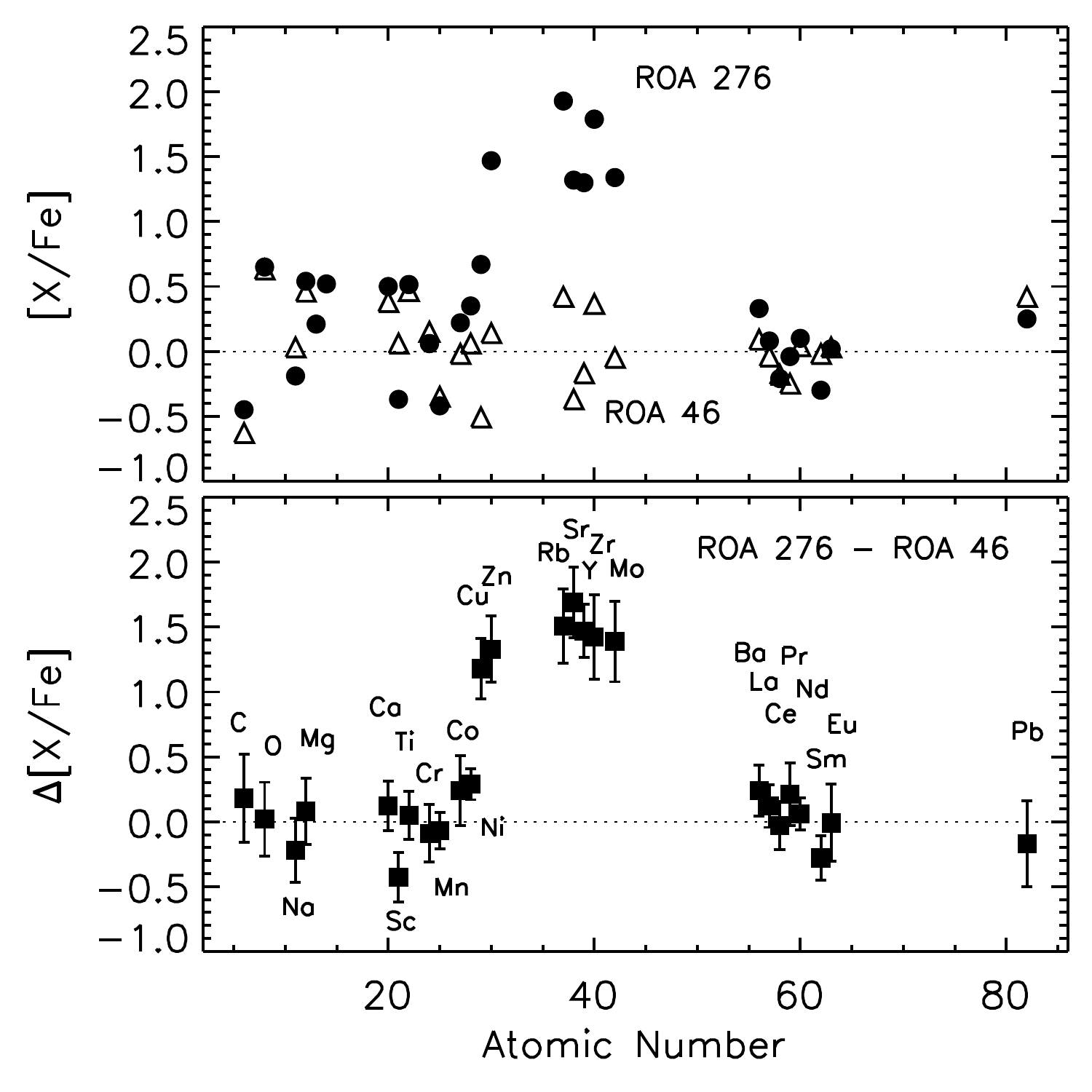}
\caption{Element abundance ratios, [X/Fe], as a function of atomic number.
(Top) Filled circles are ROA 276 and open triangles are the comparison star ROA
46. (Bottom) Relative abundance ratios in the sense ROA 276 $-$ ROA 46.} 
\label{fig:abun}
\end{figure}

We measured abundances for 28 elements in both stars, see Figure
\ref{fig:abun}. The comparison star ROA 46, with [Fe/H] = $-$1.7, belongs to
the most metal-poor primordial population of \ocen. This star has element
abundance ratios relative to iron, [X/Fe], that are typical for both field halo
stars and \ocen\ stars of comparable metallicity. For ROA 276, with [Fe/H] =
$-$1.3, the abundance ratios relative to iron for the elements from Cu to Mo 
are remarkably high and unusual, and we confirm the high [Sr/Ba] ratio reported
by \citet{Stanford:2010aa}. For other elements, the abundance ratios appear
normal when compared to stars of similar metallicity. 

In the context of the star-to-star light element abundance variations in
globular clusters \citep{Kraft:1994aa,Gratton:2004aa}, both ROA 276 and ROA 46
are ``primordial'' based on their O, Na, and Mg abundances. That is, neither
star has been affected by whatever process(es) produces the O-Na
anticorrelation in globular clusters \citep{Charbonnel:2016aa}. 

The origin of the peculiar chemical composition of ROA 276 is revealed when we
subtract the abundance pattern of the comparison star ROA 46 from ROA 276
(Figure \ref{fig:abun}, lower panel). The underlying assumptions in this
approach are that ($i$) the comparison star represents the initial, or
primordial, composition of \ocen\ and ($ii$) the peculiar star is produced by
the addition of processed material onto the primordial composition. This
methodology, of examining relative abundance ratios, has proved an extremely
effective tool for identifying the source responsible for contaminating
chemically peculiar objects \citep{Roederer:2011aa,Shingles:2014aa}. By
considering relative abundance ratios rather than absolute abundances,
systematic errors associated with a standard spectroscopic analysis largely
cancel \citep{Melendez:2009ab}. 

\section{Discussion} \label{sec:discussion}

With the above assumptions in mind, we begin the discussion by noting that the
$s$-process abundance distribution we see in stars depends upon the integrated
exposure to neutrons (a quantity usually represented by $\tau$;
\citealt{Clayton:1968aa}).  The abundance pattern we see in Figure
\ref{fig:abun} is characteristic of a low value of $\tau$, that is, a small
integrated neutron exposure, which means that elements beyond the first
$s$-process peak are difficult to synthesize (e.g., \citealt{Kappeler:2011aa}).
A low value of $\tau$ is characteristic of the $s$-process operating in
spinstars models \citep{Frischknecht:2016aa} and in intermediate-mass AGB
models (e.g., \citealt{Karakas:2012aa}), both of which release neutrons
predominantly by the $^{22}$Ne($\alpha$,$n$)$^{25}$Mg reaction. In contrast,
the $s$-process in low-mass AGB stars occurs via the
$^{13}$C($\alpha$,$n$)$^{16}$O reaction, which results in high neutron
exposures overall and invariably results in high Ba and Pb abundances when
compared to the first $s$-process peak \citep{Busso:1999aa,Karakas:2014ab}. 
We now examine model predictions from
intermediate-mass AGB stars and massive stars. 

\subsection{Intermediate-mass AGB stars}

\begin{figure}
\centering
\includegraphics[width=\columnwidth]{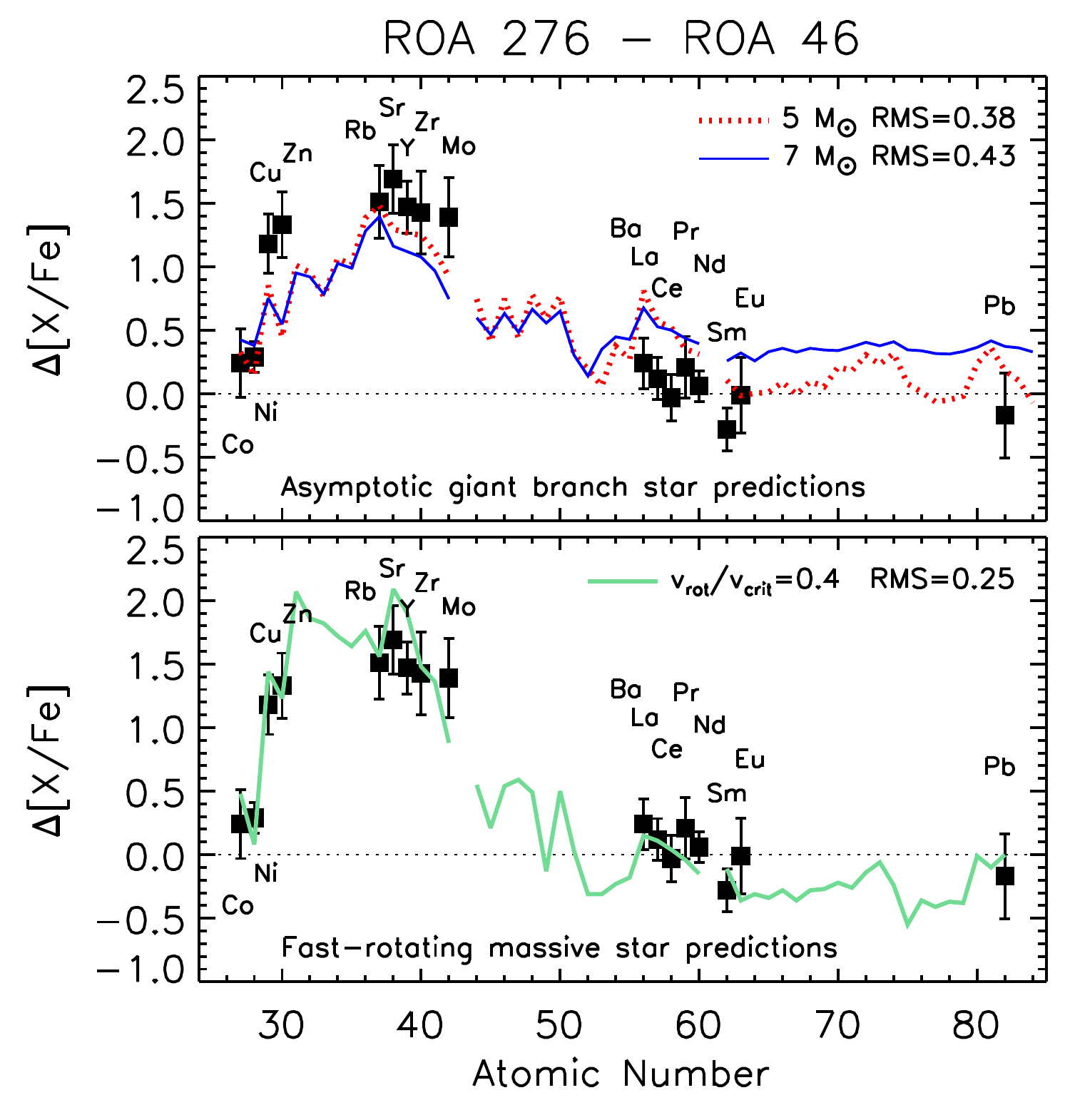}
\caption{Comparison of relative abundance ratios (ROA 276 $-$ ROA 46) and model
predictions as a function of atomic number. (Top) The dotted and solid lines
are predictions from AGB models with [Fe/H] = $-$1.2 of 5 and 7 \msun,
respectively \citep{Fishlock:2014aa}. (Bottom) The solid line is the prediction
from a 20 \msun\ stellar model with [Fe/H] = $-$1.8 rotating at 0.4 times the
critical value \citep{Frischknecht:2012aa,Frischknecht:2016aa}.} 
\label{fig:abun2}
\end{figure}

Nucleosynthesis predictions from intermediate-mass AGB models of 5 and 7 \msun\
with [Fe/H] = $-$1.2\footnote{These models adopt a scaled-solar composition.}
\citep{Fishlock:2014aa} offer an unsatisfactory fit to the data (Figure
\ref{fig:abun2}, upper panel). The RMS (root mean square) scatter between
observation and model is 0.38 dex and 0.43 dex for the 5 and 7 \msun\ models,
respectively. A similarly poor fit to the data is obtained when using AGB
yields from an independent group \citep{Cristallo:2011aa}. In particular,
nucleosynthesis occurring within non-rotating AGB models cannot simultaneously
match the high abundances of Cu and Zn along with the high ratio of light
$s$-process (e.g., Rb, Sr, Y, Zr) to heavy $s$-process elements (e.g., Ba, La,
Ce). For example, consider the following pairs of adjacent elements: ($i$) Cu
and Zn, ($ii$) Rb and Sr, and ($iii$) Ba and La. For a given pair of elements,
we compute average values, e.g., $<$Cu,Zn$>$, and ratios of these pairs, e.g.,
[$<$Cu,Zn$>$/$<$Rb,Sr$>$]. For the 5 and 7 \msun\ AGB models by
\citet{Fishlock:2014aa}, the ratios are [$<$Cu,Zn$>$/$<$Ba,La$>$] = $-$0.04 dex
and +0.05 dex, respectively, and these are a factor of 10 lower than the
observed value of +1.08 dex. A similarly large discrepancy of $\sim$0.7 dex
between the observations and AGB calculations is found for the ratio
[$<$Rb,Sr$>$/$<$Ba,La$>$]. Predictions from rotating AGB models are limited
\citep{Herwig:2003aa,Siess:2004aa,Piersanti:2013aa} and do not include detailed
$s$-process calculations for metal-poor intermediate-mass objects.  Similarly,
large grids of neutron-capture element yields for super-AGB models (rotating or
non-rotating) are unavailable. 

\subsection{Spinstars}

At low metallicity, fast rotation is an essential requirement to produce large
quantities of neutron-capture elements in massive stars
\citep{Pignatari:2008aa,Frischknecht:2012aa,Frischknecht:2016aa}. 
As noted above, the $^{22}$Ne($\alpha$,$n$) reaction is dominant in
massive stars with a negligible contribution from the $^{13}$C($\alpha$,$n$)
reaction (e.g., \citealt{Baraffe:1992aa,Nishimura:2017aa}), and the low neutron
exposure limits the production of elements beyond the first $s$-process peak
\citep{Frischknecht:2016aa}. 
Nucleosynthesis predictions from spinstars by 
\citet{Frischknecht:2012aa,Frischknecht:2016aa} provide an excellent fit to the
relative abundance ratios for all 18 elements from Cu to Pb (Figure
\ref{fig:abun2}, lower panel). At metallicity $Z$ = 10$^{-3}$, [Fe/H] =
$-$1.8\footnote{These models adopt an $\alpha$ enhancement of [$\alpha$/Fe]
$\simeq$ +0.5.}, which closely matches the comparison star ROA 46, the best fit
is obtained from the 20 \msun\ model with an initial rotation rate 0.4 times
the critical value.  The RMS is 0.25 dex, and this represents a superior fit
when compared to the AGB models.  The average measurement uncertainty is 0.24
dex, i.e., the RMS can be attributed entirely to measurement errors.
Additionally, the predicted and observed ratios for [$<$Cu,Zn$>$/$<$Ba,La$>$]
are +1.21 dex and +1.08 dex, respectively.  Agreement is also obtained for
[$<$Rb,Sr$>$/$<$Ba,La$>$] with predicted and observed values of +1.70 dex and
+1.42 dex, respectively.  Therefore, the chemical abundance pattern of ROA 276,
relative to the comparison star ROA 46, at present can be attributed to
pollution from a spinstar. It is not clear, however, whether the spinstar
polluted the gas cloud from which ROA 276 was formed or whether the pollution
occurred via binary mass transfer. 
The principal result of this work is to provide
clear observational support that the $s$-process in rapidly rotating massive
stars was a relevant nucleosynthesis source in the early Universe. 

We note that the spinstar yields of
\citet{Frischknecht:2012aa,Frischknecht:2016aa} are pre-supernova yields. The
supernova explosion does not significantly affect the pre-supernova
neutron-capture element distribution \citep{Tur:2009aa}, while major changes
are expected for other elements, like Fe and other iron-group elements
\citep{Nomoto:2013aa}. The grid of fast-rotating massive star models that we
tested covers a modest range in mass, metallicity, and rotation, and this grid
will need to be expanded in the future. We considered all the
\citet{Frischknecht:2012aa,Frischknecht:2016aa} models and adopted a threshold
RMS of 0.38 dex which corresponds to the 5 \msun\ AGB model that we regarded as
unsatisfactory. Three additional models satisfied this criterion (the RMS
values range from 0.26 dex to 0.32 dex). These models are ($i$) 25 \msun,
initial rotation rate 0.4 times the critical value, and metallicity $Z$ =
10$^{-3}$, ($ii$) same as ($i$) but with 40 \msun, and ($iii$) same as ($ii$)
but with metallicity $Z$ = 10$^{-5}$. We disregard the latter model because the
metallicity, [Fe/H] = $-$3.8, is too low compared to the program stars. At
present, the spinstar models that provide the best fits to the data have masses
between 20 and 40 \msun, rotation rate of 0.4 times the critical value, and a
metallicity of $Z$ = 10$^{-3}$. 

Spinstars, however, are predicted to synthesize large quantities of the light
elements C, N, and O \citep{Meynet:2006aa,Maeder:2015aa}. The best fitting
model predicts enhancements of $\Delta$[C/Fe] = +2.81, $\Delta$[C/Sr] = +0.72,
$\Delta$[O/Fe] = +2.99, and $\Delta$[O/Sr] = +0.90 and the observed ratios (ROA
276 $-$ ROA 46) are +0.18, $-$1.51, +0.02, and $-$1.57, respectively. The C and
O abundances in ROA 276 and ROA 46 are similar to each other and to metal-poor
field giant stars \citep{Stanford:2010aa}. Therefore, spinstar models predict
differences in C and O between ROA 276 and ROA 46 that are at least two orders
of magnitude larger than the observations. Either our proposed scenario of
pollution from spinstars is incorrect, or the current models require
refinement. 

Although we dismissed AGB stars based on the neutron-capture element abundance
distribution, here we consider the predicted yields for [C/Fe], [O/Fe], and
[C/Sr] as we did for the spinstars. The relative abundance ratios (ROA 276 $-$
ROA 46) are $\Delta$[C/Fe] = +0.18, $\Delta$[O/Fe] = +0.02, and $\Delta$[C/Sr]
= $-$1.51. The 5 and 7 \msun\ AGB models from \citet{Fishlock:2014aa} with
[Fe/H] = $-$1.2 predict [C/Fe] = +1.01 and +0.62, [O/Fe] = $-$0.20 and $-$0.64,
and [C/Sr] = $-$0.35 and $-$0.21, respectively.  Therefore, while the AGB model
predictions for C and O are not in major disagreement with the observations,
the predicted [C/Sr] ratios differ from the observations by at least an order
of magnitude. 

\subsection{ROA 46 as the comparison star} 

\begin{figure}
\centering
\includegraphics[width=2.5in]{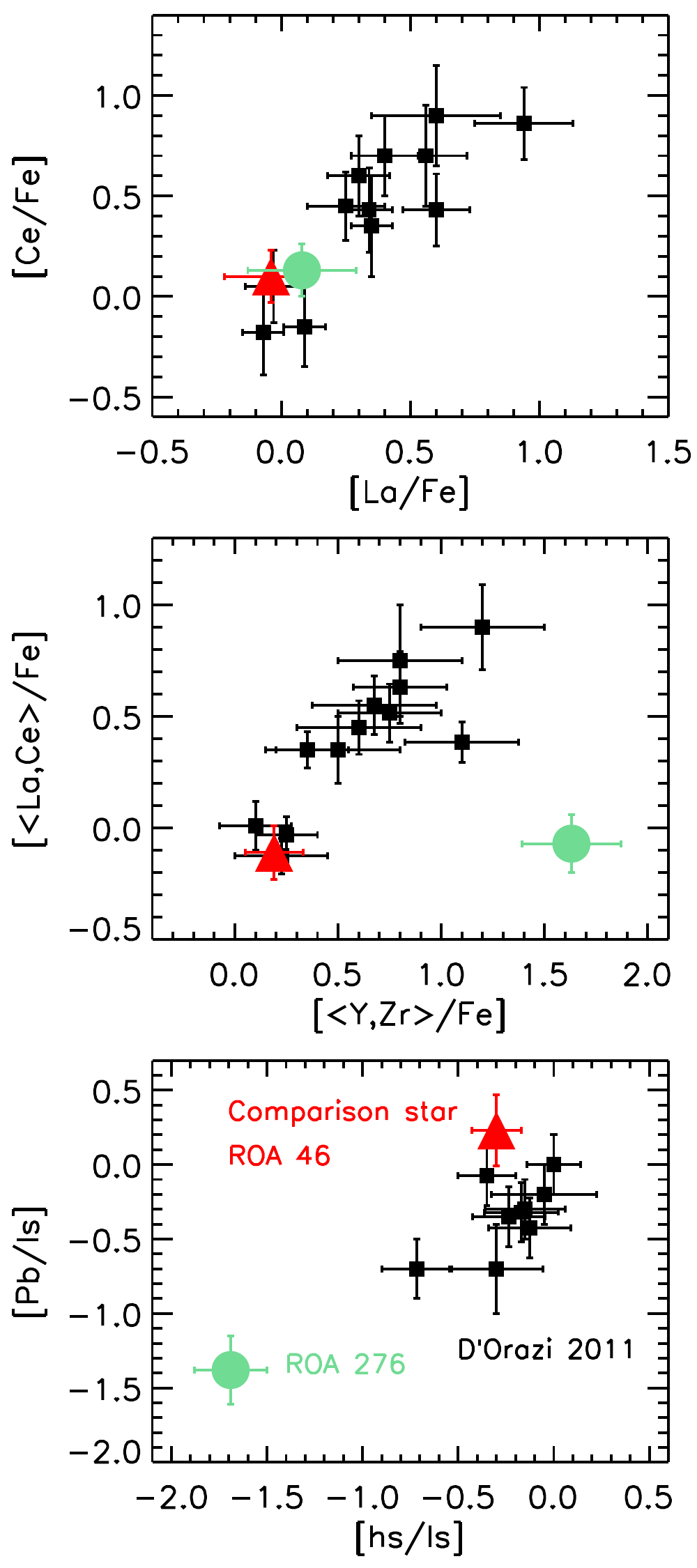}
\caption{Neutron-capture element abundance ratios for ROA 276 (aqua circle),
ROA 46 (red triangle), and \ocen\ red giants (black squares from
\citet{DOrazi:2011aa}). [Ce/Fe] vs.\ [La/Fe] is shown in the upper panel,
[$<$La,Ce$>$/Fe] vs.\ [$<$Y,Zr$>$/Fe] in the middle panel, and [Pb/ls] vs.\
[hs/ls] in the lower panel. ``hs'' is the average of La and Ce while ``ls'' is
the average of Y and Zr.} 
\label{fig:abun3}
\end{figure}

The conclusions of this work depend on the decision to use ROA46 as the
reference star. In Figure \ref{fig:abun3}, we plot various combinations of
neutron-capture element abundances and compare ROA 276 and ROA 46 with a larger
stellar sample from \ocen\ \citep{DOrazi:2011aa}. Within the measurement
uncertainties, ROA 46 is representative of the primordial population of \ocen\
with low abundance ratios for the $s$-process elements. Thus, we may consider
ROA 46 as a valid reference star. 

\subsection{Alternative nucleosynthetic sites}

Another possibility worth considering is a truncated $s$-process operation
involving the $^{13}$C($\alpha$,$n$)$^{16}$O reaction in AGB stars. For the 2
\msun\ model with [Fe/H] = $-$1.2 from \citet{Fishlock:2014aa}, we examined the
surface abundances after each thermal pulse. In contrast to ROA 276, the Ba and
Pb abundances are high relative to the first $s$-process peak elements even
after 1-2 thermal pulses and throughout the entire AGB phase. This model also
predicts large C enhancements, which is also inconsistent with the
observations. We therefore dismiss the AGB (and truncated AGB) scenario since
it does not fit the neutron-capture element distribution and also fails to
match the C and O abundances. 

Beside the $s$-process discussed in this work for spinstars and AGB stars,
other nucleosynthesis processes have been proposed as possible astrophysical
sources of heavy elements \citep{Thielemann:2011aa}. 

The rapid neutron-capture process, or $r$-process, is not consistent with the
abundance signature of ROA 276. The low C abundance and enhancements of the
light $s$-process elements in ROA 276 are qualitatively similar to the observed
abundance patterns in two halo stars (HD 122563 and HD 88609) and ET0097 in the
Sculptor dwarf galaxy \citep{Honda:2006aa,Honda:2007aa,Skuladottir:2015aa} and
are attributed to the weak $r$-process \citep{Arcones:2011aa}. The detailed
abundance pattern of the neutron-capture elements in ROA 276 (or ROA 276 $-$
ROA 46), however, does not match HD 122563, HD 88609, or ET0097.
Nucleosynthesis from the intermediate neutron-capture process or $i$-process
\citep{Cowan:1977aa,Hampel:2016aa,Jones:2016aa,Denissenkov:2017aa} is also
inconsistent with the abundance signature of ROA 276. 
 
While electron-capture supernovae are a potential source of neutron-capture
elements, at present electron-capture supernovae models do not efficiently
produce elements beyond Zr \citep{Wanajo:2011aa}. Such predictions are not
consistent with ROA 276, where a strong production is observed up to Mo. 

The alpha-rich freeze-out component ejected by high-energy core collapse
supernovae (CCSNe) does not seem to be compatible either when considering
elemental ratios in the Rb-Zr region \citep{Woosley:1992aa}. Furthermore, in
alpha-rich freeze-out conditions, it is difficult to efficiently produce
elements heavier than Zr \citep{Pignatari:2016aa}. 

Different types of neutrino-wind components in CCSNe can provide a large
scatter of abundance patterns
\citep{Frohlich:2006aa,Kratz:2008aa,Roberts:2010aa,Arcones:2011aa,Wanajo:2011ab},
and a detailed study should be undertaken to check if there are reasonable
stellar conditions that would produce material that would fit the ROA276
abundance pattern between Cu and Mo. For instance, while high-entropy wind
predictions seem to reproduce the observed pattern for Sr, Y and Zr, they
underproduce Rb \citep{Farouqi:2009aa}. 

In summary, to our knowledge there are no nucleosynthetic sources, other than
spinstars, that can match the neutron-capture element pattern in ROA 276. More
complex astrophysical scenarios involving multiple sources could be invoked to
explain the abundance pattern of ROA 276. While this would provide additional
freedom to reproduce the observed data, any such scenario might be regarded as
contrived and it is not obvious that we would find a combination of sources
that could simultaneously fit the neutron-capture element pattern without large
overabundances of C and other light elements. 

\section{Conclusions}

We present a chemical abundance analysis of the red giant ROA 276 in \ocen\ and
a comparison red giant ROA 46. The neutron-capture element distribution of ROA
276, relative to ROA 46, can be uniquely fit by nucleosynthesis predictions
from a spinstar model with 20 \msun, [Fe/H] = $-$1.8, and an initial rotation
rate 0.4 times the critical value. 

ROA 276 was originally identified from a sample of 33 red giant branch stars in
\ocen\ \citep{Stanford:2010aa}. Prior to this, examination of the Sr and Ba
abundances in 392 main sequence stars in \ocen\ revealed only one object with
high Sr and low Ba \citep{Stanford:2006ab}. Among the $\sim$1000 halo stars
with [Fe/H] $<$ $-$1 and [Sr/Ba] measurements \citep{Suda:2008aa}, only 13 have
[Sr/Ba] $>$ +1.2 and none exhibits the distinctive abundance pattern for the
suite of elements from the Fe-peak through to Pb measured in ROA 276. Objects
with chemical compositions similar to ROA 276 are rare. The predicted [Sr/Ba]
ratio from spinstars varies with mass and metallicity, and very high [Sr/Ba]
ratios only occur around metallicities [Fe/H] = $-$2 to $-$1. ROA 276 (perhaps
thanks to its metallicity and environment) provides a unique stellar laboratory
to study neutron-capture nucleosynthesis in spinstars. 

\bigskip
\acknowledgements

We thank the referee for helpful comments.
D.Y., J.E.N., G.D.C, and A.I.K.\ acknowledge support from the Australian
Research Council (grants DP0984924, FT110100475, DP120100475, DP120101237,
FT140100554, and DP150103294). R.H.\ acknowledges support from
EU-FP7-ERC-2012-St Grant 306901. A.I.K.\ and R.H.\ acknowledge support from the
World Premier International Research Center Initiative (WPI Initiative), MEXT,
Japan. MP acknowledges significant support to NuGrid from NSF grants PHY
09-22648  (Joint Institute for Nuclear Astrophysics, JINA), NSF grant
PHY-1430152  (JINA Center for the Evolution of the Elements) and EU
MIRG-CT-2006-046520. MP acknowledges the support from the "Lend{\" u}let-2014"
Programme of the Hungarian Academy of Sciences (Hungary) and from SNF
(Switzerland). 
Australian access to the Magellan Telescopes was supported through the Major
National Research Facilities program of the Australian Federal Government. 
We acknowledge financial support from the Access to Major Research Facilities
Program, under the International Science Linkages Program of the Australian
Federal Government. 

\facilities{Magellan:Clay (MIKE)}

\bibliographystyle{apj}

\end{document}